\documentclass[a4paper,12pt,twoside]{article} 
\usepackage{a4wide}
\usepackage{rotating}

\usepackage{graphicx}                
\usepackage{amssymb}

\newcommand\jhep[3] {{{\it J. High Energy Phys.\ }{\bf #1} (#2) #3}}
\newcommand\npb[3] {{{\it Nucl.\ Phys.\ }{\bf B #1} (#2) #3}}
\newcommand\npps[3] {{{\it Nucl.\ Phys.\ }{\bf #1} {\it(Proc.\ Suppl.)} (#2) #3}}
\newcommand\plb[3] {{{\it Phys.\ Lett.\ }{\bf B #1} (#2) #3}}
\newcommand\prd[3] {{{\it Phys.\ Rev.\ }{\bf D #1} (#2) #3}}
\newcommand\prl[3] {{{\it Phys.\ Rev.\ Lett.\ }{\bf #1} (#2) #3}}
\newcommand\rmp[3] {{{\it Rev.\ Mod.\ Phys.\ }{\bf #1} (#2) #3}}
\newcommand\sjnp[3] {{{\it Sov.\ J.\ Nucl.\ Phys.\ }{\bf #1} (#2) #3}}

\begin{document}

{\flushright

FTUV-020109

}

\begin{center}
\noindent{\Large \tt \bf 

The sign of $\Delta m^{2}_{31}$ and the muon-charge asymmetry for
atmospheric neutrinos
 
}\vspace{10mm}
\renewcommand{\thefootnote}{\fnsymbol{footnote}}
\noindent{\large
J. Bernab\'eu\footnote{Talk given at the 7th International Workshop on Topics in Astroparticles and Underground Physics (TAUP2001), Laboratori Nazionali del Gran Sasso, Italy, September 8-12, 2001.} and
S. Palomares-Ruiz 
}\vspace{6mm}

\noindent{\it
  Departament de F\'{\i}sica Te\`orica 

  Universitat de Val\`encia  
      
  46100 Burjassot, Val\`encia, Spain
\\
}
\end{center}
\vspace{6mm}
\renewcommand{\thefootnote}{\arabic{footnote}}
\setcounter{footnote}{0}

\begin{abstract}We discuss the possibility to measure the sign of $\Delta 
m^{2}_{31}$ from matter-induced charge asymmetries in atmospheric
neutrino oscillations. The main conclusion is that an impact on the
$\nu_{\mu}$ survival probability requires the action of the MSW
resonance, which becomes visible for baselines above $\sim$ 7000 km.
\end{abstract}

\newpage

  \section{Introduction}

Present evidence for neutrino masses and mixings can be summarized as:
1) the atmospheric $|\Delta m^{2}_{31}| \sim (1-5) \cdot 10^{-3}
$eV$^2$ is associated with a mixing, $\theta_{23}$, near to maximal
\cite{SK}; 2) the solar $\Delta m^{2}_{21}$ prefers the LMA-MSW
solution \cite{solar}; CHOOZ reactor data \cite{chooz} give severe
limits for $|U_{e 3}|$. In this contribution we are going to discuss
that contrary to a wide spread belief, Earth effects on the
propagation of atmospheric neutrinos can become observable
\cite{earth} even if $|U_{e 3}|$ is small, but non-vanishing. This
fact would allow to determine the sign of $\Delta m^{2}_{31}$
\cite{barger}. For baselines $L$ smaller than the Earth diameter,
appropiate for atmospheric neutrinos, $\frac{\Delta m^{2}_{21}}{4 E} L
\equiv \Delta_{21} \ll 1$, so that we will neglect the
(1,2)-oscillating phase in vacuum against the (2,3)-one. This is a
very good aproximation, unless the high $\Delta m^{2}_{21}$-region of
the LMA solution turns out to be the solution to the solar problem. In
that case we should take into account corrections of order $O
(\frac{\Delta m^{2}_{21}}{\Delta m^{2}_{31}})$ (see
eg. \cite{corrections}).

In section 2 we discuss the correspondence between the determination
of the sign of $\Delta m^{2}_{31}$ and the observation of the Earth
effects in a transition involving $\nu_e$. The change expected in the
neutrino spectrum and mixing due to matter effects is pointed
out. Section 3 studies the observability of the MSW-resonance, with a
positive conclusion for baselines $L \gtrsim$ 7000 km, and its impact
on the survival probability, $\nu_{\mu} \rightarrow
\nu_{\mu}$. Section 4 gives an analysis of the matter-induced CPT-odd
asymmetry, together with the realistic charge-asymmetry expected for
atmospheric neutrinos. In section 5 we present some conclusions.

  \section{The neutrino spectrum in matter}

Current analyses leave us with two alternatives for the spectrum of
the three active neutrino species, either hierarchical or degenerate.

The effective neutrino potential due to the charged current
interaction of $\nu_e$ with the electrons in the medium is \cite{wolf}
$V \equiv \frac{a}{2 E} = \sqrt{2} G_F N_e$, so that the effective
hamiltonian, in the extreme relativistic limit, is given by
\cite{kuo-pan}

\begin{equation}
\label{h} 
H = \frac{1}{2 E} \left\{ U \left( \begin{array}{ccc} 0 & \; 0 & 0 \\
0 & \; 0 & 0 \\ 0 & \; 0 & \Delta m^{2}_{31} \end{array} \right)
U^{\dagger} + \left( \begin{array}{ccc} a & 0 & 0 \\ 0 & 0 & 0 \\ 0 &
0 & 0 \end{array} \right) \right\}
\end{equation}

In going from $\nu$ to $\overline{\nu}$, there are matter-induced CP-
and CPT- odd effects associated with the change $a \rightarrow -
a$. The additional change U $\rightarrow$ U$^*$ is inoperative in the
limit of (\ref{h}). The effects we are going to discuss depend on the
interference between the different flavors and on the relative sign
between $a$ and $\Delta m^{2}_{31}$. As a consequence, an experimental
distinction between the propagation of $\nu$ and $\overline{\nu}$ (the
sign of $a$) will determine the sign of $\Delta m^{2}_{31}$. An
appreciable interference will be present if and only if there are
appreciable matter effects. For atmospheric neutrinos, one needs the
``connecting'' mixing $U_{e 3}$ between the $\nu_e$-flavor and the
$\nu_3$ mass eigenstate to show up.

For small $s_{13}$ \cite{chooz}, even if the effects on the spectrum
are expected to be small with respect to the decoupling of the
$\nu_e$-flavour in matter, there could be a substantial mixing of
$\nu_e$ with $\tilde{\nu}_3$ if one is near to a situation of
level-crossing. This would lead to a resonant MSW behaviour
\cite{msw}.

\begin{equation}
\label{resonance}
\sin^2{2 \, \tilde{\theta}_{13}} = \frac{4 \, s^{2}_{13} \, c^{2}_{13}}{(\alpha
- \cos{2 \, \theta_{13}})^2 + 4 \, s^{2}_{13} \, c^{2}_{13}} \hspace{0.3cm},
\hspace{0.7cm} \alpha \equiv \frac{a}{\Delta m^{2}_{31}}
\end{equation}

But still $\langle \tilde{\nu}_1 | \nu_e \rangle = 0$, i.\ e., the
$\nu_e$ has no overlap with the lowest mass eigenstate in matter. This
vanishing mixing in matter is responsible for the absence of
fundamental CP-violating effects, even if there are three
non-degenerate mass eigenstates in matter. In vacuum, the absence of
genuine CP-odd probabilities was due to the degeneracy $\Delta_{21} =
0$. The step from vanishing $\Delta_{21}$ in vacuum to the vanishing
mixing $U_{e 1}$ in matter was termed a ``transmutation''
\cite{trans}.

  \section{Observability of the MSW resonance}

For atmospheric $\nu_{\mu}$ neutrinos, matter effects in the survival
probability $\nu_{\mu} \rightarrow \nu_{\mu}$ would be minute unless
the resonance shows up. The resonance is not apparent even at $L$ =
3000 km, appropiate for neutrino factories \cite{nuf}. Is there a way
out?

Again, a non-vanishing connecting mixing $s_{13} \neq 0$ provides the
solution. Along with it, there is a resonance width which, when
discussed in terms of the dimensionless parameter $\alpha$, is given
by

\begin{equation}
\label{alpha}
\alpha_R = \cos{2 \, \theta_{13}} \hspace{0.3cm}, \hspace{0.7cm}
\Gamma_{\alpha} = 2 \; \sin{2 \, \theta_{13}}
\end{equation}

One discovers that the oscillating phase on the resonance is
non-vanishing, but given by the $L$-dependent relation

\begin{equation}
\label{res}
\tilde{\Delta}_{31 (R)} = \Delta_{31} \; \frac{\Gamma_{\alpha}}{2}
\end{equation}
  
If $L \ll L_{opt}$, with optimal $L$, $L_{opt}$, defined by
$\tilde{\Delta}_{31 (R)} = \pi /2$, the resonance does not affect the
oscillation probability. On the contrary, around $L_{opt} = \frac{2 \;
\pi}{\tilde{a} \; \tan{2 \, \theta_{13}}}$, where $\tilde{a} = a/E$,
the resonance becomes apparent and $L_{opt}$ is independent of $\Delta
m^{2}_{31}$, which determines the resonant energy. For $L = L_{opt}$,
the maximum mixing is accompanied by maximum oscillating factor.

Under these conditions, all channels would see the resonant effect. Contrary to
non-resonant matter effects, the resonance only affects the (anti)neutrino
channels if $\Delta m^{2}_{31} > 0 (< 0)$.

  \section{Charge asymmetries}

As discussed in section 3, matter effects distinguish neutrinos from
antineutrinos. It is convenient to present them in terms of CP-odd
(for appearance channels) and CPT-odd (for the survival probabilities)
asymmetries. In the limit $\Delta_{21} = 0$, there is no room for
genuine CP violation. The interaction with matter will generate an
asymmetry effect, however, which is not connected with the vacuum
propagation.

\noindent 
For $\nu_{\mu}$ and $\bar{\nu}_{\mu}$, one has

\begin{figure}[t]
\begin{center}
\includegraphics[width=10cm]{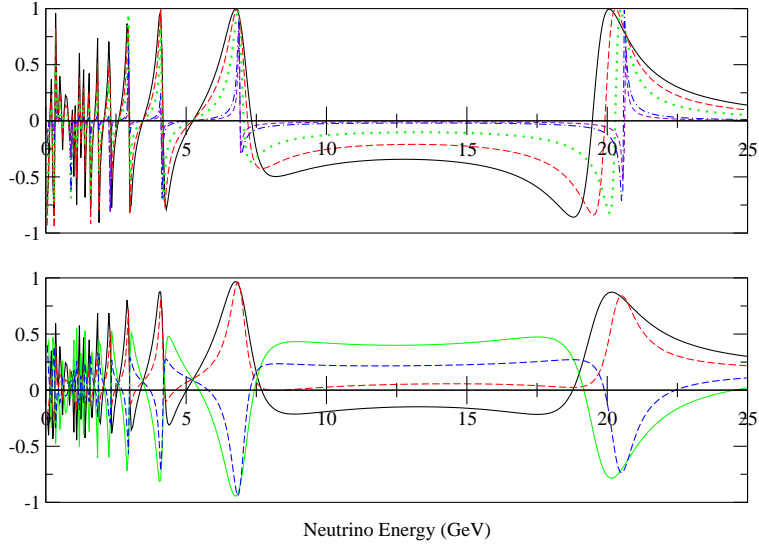} 
\caption{Upper panel: CPT-asymmetry, $A_{CPT}$, for different
values of $\sin^{2}{2 \theta_{13}}$. From up to down: $\sin^{2}{2
\theta_{13}}$ = 0.005, 0.01, 0.05, 0.10, 0.16 and $\Delta m^{2}_{31} >
0$ as all the plots are symmetric with respect to the horizontal axis
when $\Delta m^{2}_{31} < 0$. Lower panel: Charge-asymmetry, A, for
$\sin^{2}{2 \theta_{13}}$ = 0.05 (dashed line) and 0.16 (solid
line). The lower plots correspond to $\Delta m^{2}_{31} > 0$ and the
upper ones to $\Delta m^{2}_{31} < 0$. For both panels, $L = 8000$ km,
$\sin^{2}{2 \theta_{23}}$ = 1 and $\left|\Delta m^{2}_{31}\right| =
3.2 \cdot 10^{-3}$ eV$^2$.  \label{asym}}
\end{center}
\end{figure}

\begin{equation}
\label{acpt}
A_{CPT} = \frac{P (\nu_{\mu} \rightarrow \nu_{\mu} ; L) - 
P (\bar{\nu}_{\mu} \rightarrow \bar{\nu}_{\mu} ; L)}
{P (\nu_{\mu} \rightarrow \nu_{\mu} ; L) + 
P (\bar{\nu}_{\mu} \rightarrow \bar{\nu}_{\mu} ; L)}
\end{equation}

\noindent
and it is represented in fig. \ref{asym} as function of the energy for
a baseline of $L = 8000$ km and different values of $\sin^2{2 \,
\theta_{13}}$.  Around the resonance, $A_{CPT}$ presents a plateau
with non-vanishing appreciable values (depending on $\sin^2{2 \,
\theta_{13}}$). The big asymmetries at 6 and 20 GeV correspond to low
probabilities and they are not of interest. The negative (positive)
asymmetry in the plateau is obtained for $\Delta m^{2}_{31} > 0 (<
0)$. Obviously, it is symmetric with respect to the horizontal axis
when changing the sign of $\Delta m^{2}_{31}$. As we have seen above,
the optimal baseline is inversely proportional to the $\theta_{13}$
mixing.

For atmospheric neutrinos, $A_{CPT}$ cannot be separated out and the $\nu_e
(\overline{\nu}_e)$ flux also contributes to the detection of $\nu_{\mu}
(\overline{\nu}_{\mu})$. Taking into account the CC cross-sections in the
detector,

\begin{equation}
\label{nevents}
\begin{array}{l}
N(\mu^-;E) = \sigma_{cc}(\nu_{\mu}) \; \left[ \phi^o (\nu_{\mu};E) \;
P(\nu_{\mu} \rightarrow \nu_{\mu}) + \phi^o (\nu_e:E) \; P(\nu_e
\rightarrow \nu_{\mu}) \right] \\[2ex] N(\mu^+;E) =
\sigma_{cc}(\overline{\nu}_{\mu}) \; \left[ \phi^o
(\overline{\nu}_{\mu};E) \; P(\overline{\nu}_{\mu} \rightarrow
\overline{\nu}_{\mu}) + \phi^o (\overline{\nu}_e;E) \;
P(\overline{\nu}_e \rightarrow \overline{\nu}_{\mu}) \right]
\end{array}
\end{equation}   

\noindent
where $\phi^o (\nu_{\mu};E)$ ($\phi^o (\overline{\nu}_{\mu};E)$) and
$\phi^o (\nu_e;E)$ ($\phi^o (\overline{\nu}_e;E)$) are the muon and
electron (anti) neutrino fluxes, respectively, calculated from
\cite{fnv}. As in the important energy range, both cross-sections are,
to good aproximation, linear with the energy, one can build an
asymmetry which eliminates what is induced by $\sigma_{cc}$ in the
form

\begin{equation}
\label{a}
A = \frac{N(\mu^-;E) - \frac{\sigma_{cc}(\nu_{\mu})}
{\sigma_{cc}(\overline{\nu}_{\mu})} \; N(\mu^+;E)}{N(\mu^-;E) +
\frac{\sigma_{cc}(\nu_{\mu})}{\sigma_{cc}(\overline{\nu}_{\mu})} \;
N(\mu^+;E)}
\end{equation}

In (\ref{a}) there is still some asymmetry generated by the atmospheric
neutrino fluxes. Contrary to $A_{CPT}$, the value of the muon-charge asymmetry
is not symmetric with respect to the abscisa axis when changing the sign of
$\Delta m^{2}_{31}$. In fig. \ref{asym} we give the values of A for two values
of $\sin^2{2 \, \theta_{13}}$. There is again an appreciable separation between
the cases of positive and negative $\Delta m^{2}_{31}$.

  \section{Conclusions}

In the limit of $\frac{\Delta m^{2}_{21}}{4 E} L \ll 1$, the main
conclusions of this study are: i) The medium effects, which
discriminate between neutrino and antineutrino propagation determine
the sign of the atmospheric $\Delta m^{2}_{31}$; ii) for $s_{13} = 0$,
electron neutrinos decouple from neutrino mixing in matter and have a
definite effective mass in matter; iii) for $s_{13} \neq 0$, electron
neutrinos mix with the third mass eigenstate neutrino and take part in
the atmospheric neutrino oscillations; iv) electron neutrinos do not
mix with the first mass eigenstate in matter, avoiding the generation
of genuine CP-violating effects; v) non-resonant medium effects are
already apparent in the sub-sominant channel $\nu_e \rightarrow
\nu_{\mu}$ for baselines $L \sim 3000$ km, in both the mixing and the
oscillation phase-shift; vi) the observation of matter effects in the
$\nu_{\mu}$-survival probability requires the action of the MSW
resonance, with baselines longer than $L \sim 7000$ km; vii) the
optimal baseline depends on the value of $s_{13}$, but the effects are
much cleaner in the region of the longest baselines without entering
the Earth core \cite{earth} (nadir angles $\theta_n \gtrsim
33^o$). The corresponding muon-charge asymmetry shows an appreciable
separation for the two possible signs of $\Delta m^{2}_{31}$.\\

This work is supported by Grant AEN-99-0296.

\end{document}